\documentclass[aps,prb,twocolumn,groupedaddress]{revtex4}
\usepackage[dvips]{epsfig}
\usepackage{amsmath}

\begin{document}

\title{Magnetoelastic coupling in triangular lattice antiferromagnet CuCrS$_2$}

\author{Julia C. E. Rasch$^{1,2}$}
\email[]{rasch@ill.fr}
\author{Martin Boehm$^1$}
\author{Clemens Ritter$^1$}
\author{Hannu Mutka$^1$}
\author{J\"urg Schefer$^2$}
\author{Lukas Keller$^2$}
\author{Galina M. Abramova$^3$}
\author{Antonio Cervellino$^4$}
\author{J\"org F. L\"offler$^5$}
\affiliation{$^1$ Institut Laue-Langevin, 6 Rue Jules Horowitz, BP 156, 38042 Grenoble Cedex 9, France\\
$^2$ Laboratory for Neutron Scattering, ETH Zurich \& Paul Scherrer Institut, CH-5232 Villigen, PSI, Switzerland\\
$^3$ L.V. Kirensky Institute of Physics, SB RAS, Krasnoyarsk 660036, Russia\\
$^4$ Swiss Light Source, Paul Scherrer Institut, CH-5232 Villigen, PSI, Switzerland\\
$^5$ Laboratory of Metal Physics and Technology, Department of Materials, ETH Zurich, 8093 Zurich, Switzerland}

\date{\today}

\begin{abstract}
CuCrS$_2$ is a triangular lattice Heisenberg antiferromagnet with a rhombohedral crystal structure. We report on neutron and synchrotron powder 
diffraction results which reveal a monoclinic lattice distortion at the magnetic transition and verify a magnetoelastic coupling. CuCrS$_2$ is 
therefore an interesting material to study the influence of magnetism on the relief of geometrical frustration.
\end{abstract}

\maketitle

\section{Introduction}

\begin{figure}
\includegraphics[width=\columnwidth]{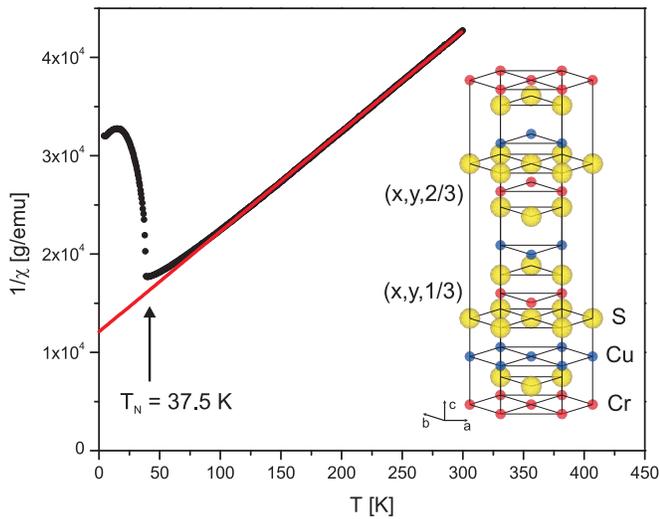}
\caption{(Color online) Magnetic susceptibility of CuCrS$_2$ measured at $H=10$~kOe as a function of temperature. An antiferromagnetic anomaly at 
$T_N=37.5$~K is clearly visible. The red solid line shows the Curie-Weiss fit to the paramagnetic part of the curve. Inset: Structural unit cell of 
CuCrS$_2$ at room temperature with space group $R3m$.}
\label{unitcell}
\end{figure}

Heisenberg antiferromagnets on a triangular lattice are subject to ongoing interest due to a strong correlation between lattice geometry, electronic 
and magnetic properties. Geometrical frustration plays a key role in compounds with triangular arrangement of magnetic moments \cite{Collins1997} and 
leads to particular characteristics including incommensurate spin structures and multiferroic properties.\cite{Cheong2007} Ternary triangular lattice 
dioxides and dichalcogenides have in common to crystallize in a layered structure with strong crystalline anisotropy perpendicular to the layers 
(mainly space group $R3m$ and $R\bar3m$). In chromium based dioxides the most commonly established magnetic structure is a quasi two-dimensional 
120$^{\circ}$ spin structure, typical for Heisenberg exchange on a triangular lattice, with a weak interlayer coupling e.g. in AgCrO$_2$ 
\cite{Oohara1994}, PdCrO$_2$ \cite{Takatsu2009}, NaCrO$_2$ \cite{Hsieh2008} and CuCrO$_2$ \cite{Kadowaki1990}. The latter two also show a 
magnetoelectric coupling where the spin structure induces ferroelectricity \cite{Seki2008,Kimura2008}. Ternary chromium dichalcogenides have been less 
intensively studied, and show a variety of different magnetic structures ranging from a 120$^{\circ}$ spin structure in LiCrS$_2$ \cite{Lafond2001} 
over a commensurate magnetic structure in KCrS$_2$ \cite{Engelsman1973} to a helix with an in-plane spin orientation in 
NaCrS$_2$.\cite{Engelsman1973}\\
The compound under investigation, CuCrS$_2$, has drawn attention as an ionic conductor \cite{Engelsman1973} and is assumed to exhibit a spin glass 
state under the substitution of chromium ions by vanadium.\cite{Abramova2005,Tsujii2007} A microscopic temperature-dependent investigation of magnetic 
and crystallographic properties is, however, lacking. CuCrS$_2$ crystallizes in space group $R3m$ at room temperature.\cite{Wilson1969} The basic 
atomic structure consists of covalent-ionic bound S-Cr-S layers separated by a van der Waals gap causing strong crystalline anisotropy perpendicular 
to the layers. Isolated CrS$_2$ is metastable and occurs only with an electron donor, therefore monovalent Cu$^+$ cations are intercalated between 
(CrS$_2$)-sandwiches.\cite{Engelsman1973,Bongers1968} The magnetic Cr$^{3+}$ ions with spin $S=3/2$ form triangular layers which are shifted by 
$(\frac{1}{3},\frac{2}{3},\frac{2}{3})$ and $(\frac{2}{3},\frac{1}{3},\frac{1}{3})$, as shown in the inset of Fig.~\ref{unitcell}. The 
nearest-neighbor intralayer Cr-Cr distance of $3.48$~\AA\ is small compared to the interlayer Cr-Cr distance of $6.55$~\AA.\ The magnetic exchange 
within the layers over Cr-S-Cr pathways is expected to be much stronger than the interlayer exchange via Cr-S-Cu-S-Cr bonds. Hence, one would expect 
to find the crystalline anisotropy to be reflected in the magnetic system, as described in the dioxides. Nevertheless, early neutron powder 
diffraction experiments on CuCrS$_2$ revealed a three-dimensional magnetic ordering below $T_N=40$~K into a complex helical structure with an 
incommensurate magnetic propagation vector.\cite{Wintenberger1987} This phenomenon cannot be explained without interlayer exchange interactions which 
must be of the same order of magnitude as the intralayer exchange. An estimation of exchange constants is given in the discussion. The particularity 
of CuCrS$_2$ is a strong lattice effect exactly at $T_N$. We assume that the lattice distortion provides a channel for the system to escape the 2D 
triangular arrangement, usually leading to magnetic frustration. The consequences on the lattice and magnetic structure throughout the phase 
transition are subject of this paper. A similar lattice effect occurs in triangular lattice CuFeO$_2$, where a crystal symmetry lowering is found to 
lift the degeneracy of the frustrated spin system \cite{Terada2006} together with multiferroic properties induced by a magnetic field or non magnetic 
impurities.\cite{Seki2007,Kimura2006} Strong spin-lattice effects are also found in 3D frustrated systems, as shown for the well-known pyrochlore 
spinel ZnCr$_2$O$_4$.\cite{Lee2000} The compound was found to undergo a Spin-Peierls-like phase transition which comprises an energy lowering through 
a lattice distortion and the opening of an energy gap. The origin of this gap has been explained by clustering of the Cr moments into weakly 
interacting antiferromagnetic loops.\cite{Lee2002} Although the formation of spin cluster has been clarified, the magnetic structure is not fully 
understood.\cite{Glazkov2009} We claim to see a similar effect on a triangular lattice and report in this paper on the detailed magnetic structure.

\section{Experimental Details}

The polycrystalline sample of CuCrS$_2$ was synthesized by solid state reaction of the pure elements. A detailed description is given elsewhere 
\cite{Almukhametov2003}. Single crystals were grown by chemical vapor transport in an evacuated quartz ampoule with $p\le10^{-4}$~mbar. A temperature 
gradient was applied over 30~cm ampoule length between 900$^{\circ}$C and 950$^{\circ}$C during three weeks. The furnace was then cooled at a cooling 
rate of 1~K/min to room temperature. The resulting single crystals are thin platelets which possess a black shiny surface with the [001] direction 
perpendicular to it. The size of the surface ranges from 10 to 100 mm$^2$ with a typical thickness of 0.2~mm. The stoichiometry of the samples was 
checked by energy dispersive X-ray spectroscopy and verified in case of the polycrystalline powder. The single crystals contain an impurity phase of 
not more than 10\% CuCr$_2$S$_4$. Systematic variations of growth conditions in order to avoid the impurity phase, produced crystals too thin 
($<10$~$\mu$m) and unstable for use in neutron scattering experiments.\\
Neutron powder diffraction measurements were carried out on the cold diffractometer DMC at the neutron spallation source SINQ, Paul Scherrer Institut 
(PSI) and the thermal diffractometer D1A at the Institut Laue-Langevin (ILL) with $\lambda=2.46$~\AA\ and $\lambda=1.91$~\AA\ , respectively, in a 
standard orange cryostat. Single crystal diffraction experiments were performed on the thermal instrument TriCS \cite{Schefer2000} (SINQ) with 
$\lambda=1.18$~\AA\ in a four-circle Eulerian cradle and on the three-axis spectrometer IN3 (ILL) with $\lambda=2.36$~\AA. High-resolution synchrotron 
radiation powder diffraction patterns were collected at the powder diffraction station of the Swiss Light Source Materials Science (SLS-MS) beamline 
at an incident photon energy of $\lambda=0.621288(1)$~\AA. The sample was mounted in a Janis cryostat in 0.2~mm Lindemann capillaries which spun at 
approximately 10~Hz during the 2$\theta$ scan to avoid preferred orientation. The collected neutron and X-ray diffraction data were refined using the 
program Fullprof \cite{Carvajal1993} based on Rietveld refinement.\\
Magnetic susceptibility measurements were carried out on a physical properties measurement system over a temperature range of $4-300$~K with an 
external field of 10~kOe and on a SQUID magnetometer for temperatures up to 400~K.

\section{Results}

\begin{figure}
\includegraphics[width=\columnwidth]{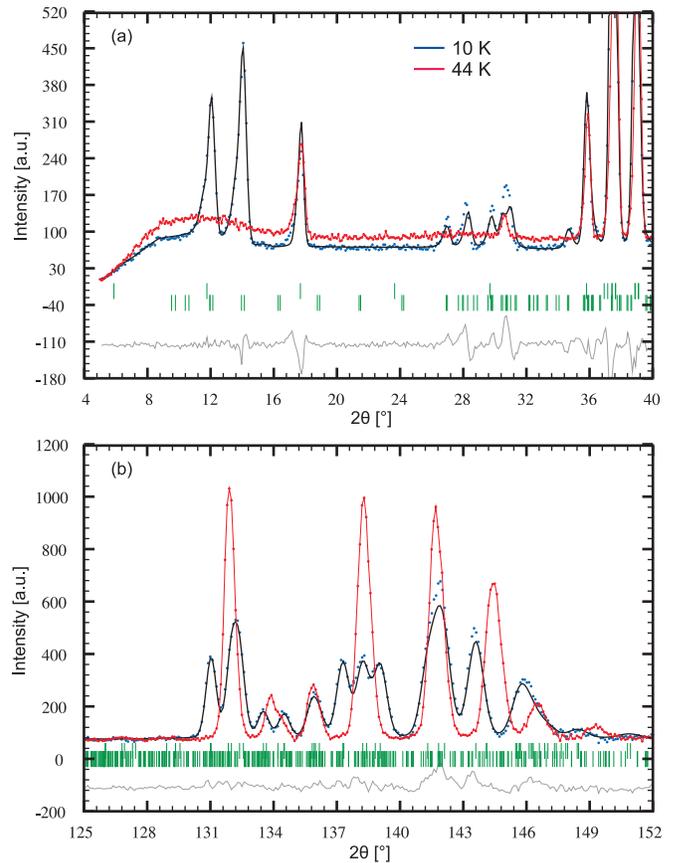}
\caption{(Color online) Powder diffraction data taken on D1A at 10~K (refined) and 44~K (raw data). The refinement of the 10~K data with a nuclear and 
magnetic phase is indicated by a black solid line. The difference between refinement and observed data at 10~K is shown in grey. (a) depicts the low 
$2\theta$ region where magnetic reflections are visible which disappear at 44~K and (b) shows the high $2\theta$ region where a splitting of the 
nuclear Bragg peaks below the transition is observable.}
\label{D1A}
\end{figure}
\begin{figure}
\includegraphics[width=\columnwidth]{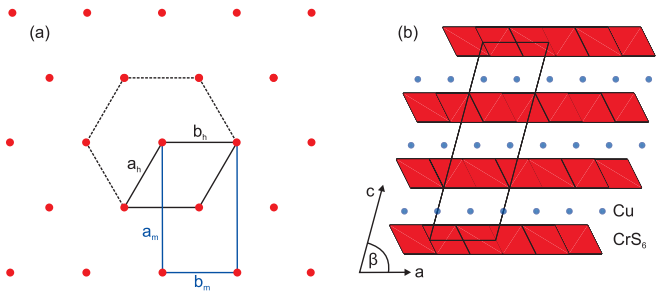}
\caption{(Color online) (a) Correlation of the rhombohedral unit cell in hexagonal description ($R3m$) and the monoclinic unit cell ($Cm$). (b) 
Influence of the monoclinic angle $\beta$ on the layers. The angle is overdrawn to show the effect of distortion; actually $\beta$ only deviates 
slightly from 90$^{\circ}$.}
\label{monoclinic}
\end{figure}
\begin{figure}
\includegraphics[width=\columnwidth]{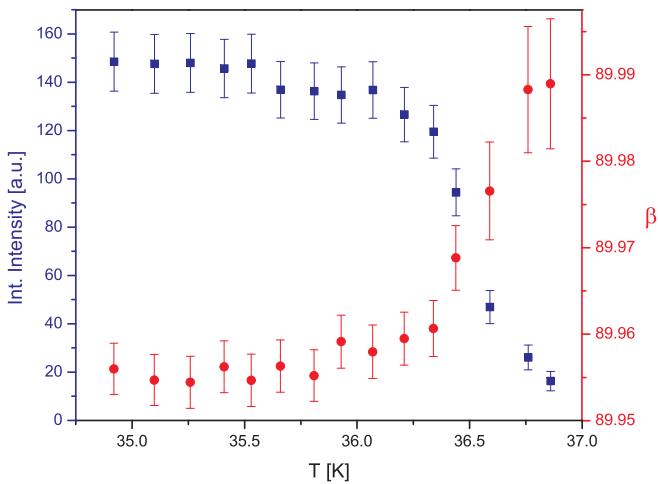}
\caption{(Color online) Intensity of the strongest magnetic Bragg peak at $2\theta=13.96^{\circ}$ (squares) and the monoclinic angle $\beta$ (circles) 
as a function of temperature, extracted from neutron powder diffraction data taken on D1A. While cooling the sample, the ordering of magnetic moments 
and the lattice distortion set in simultaneously.}
\label{angle}
\end{figure}
\begin{figure}
\includegraphics[width=\columnwidth]{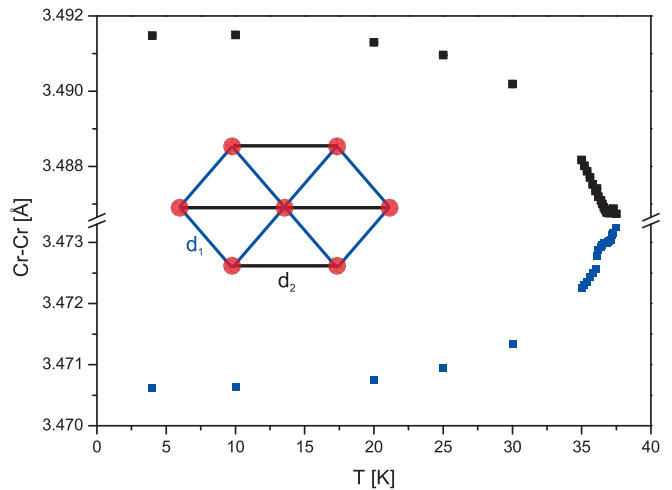}
\caption{(Color online) Temperature dependence of in-plane Cr-Cr bonds measured by synchrotron powder diffraction. Inset: Different kinds of chromium 
bonds and their influence on the distortion. The deformation is overdrawn for better illustration.}
\label{hexagon}
\end{figure}

Figure \ref{unitcell} shows the temperature dependence of the magnetic susceptibility of the powder sample with an antiferromagnetic transition at 
$T_N=37.5$~K. From the linear part of the curve the paramagnetic Curie-Weiss temperature could be determined to $\Theta_{CW}=-118.5(1)$~K and the 
effective magnetic moment amounts to $\mu_{exp}=3.75(1)$~$\mu_B$ close to the spin-only value $\mu_{S_{3/2}}=3.87$~$\mu_B$. The magnetisation curve of 
the single crystal, measured up to 400~K (not shown), indicates a superposition of the antiferromagnetic transition of CuCrS$_2$ at $T_N=37.5$~K and a 
ferromagnetic transition of CuCr$_2$S$_4$ at $T_c=377$~K.\cite{Kamihara2004}\\
A good measure for the degree of frustration is the ratio $|\Theta_{CW}|/T_N$ which is in case of CuCrS$_2$ only 3.16 and therefore rather small to 
infer a strongly frustrated system with values usually larger than 10.\cite{Greedan2000}\\
Neutron powder diffraction data taken on D1A and DMC at different temperatures show strong magnetic Bragg peaks emerging below the magnetic 
transition, i.e. $T_N=37.5$~K. This observation must be related to a magnetic ordering of the system. At higher diffraction angles $2\theta$, 
accessible on D1A, a splitting of nuclear Bragg peaks is also observable, which indicates a symmetry lowering of the crystal structure. The pattern 
taken above and below the transition at 44~K and 10~K, respectively, are shown in Fig.~\ref{D1A}. The pictures (a) and (b) show the refinement of the 
low temperature data at 10~K with two phases, nuclear and magnetic, superimposed by the raw data pattern at 44~K. Above $T_N$ the Cr$^{3+}$ moments 
show no long range order anymore, but localized Cr$^{3+}$ moments still cause paramagnetic scattering. Therefore, the difference in background 
intensity between spectra taken at 10~K and 44~K at low angles $2\theta$ (Fig.~\ref{D1A}(a)) can be explained by an angle dependence of the 
paramagnetic scattering due to the magnetic form factor. The effect is negligible at higher angles $2\theta$ (Fig.~\ref{D1A}(b)).\\
The structural transition indicated by a splitting of the nuclear Bragg peaks in the low temperature phase could be identified as a symmetry lowering 
from rhombohedral space group $R3m$ (160) to monoclinic space group $Cm$ (8). The relation between both unit cells is given in Fig.~\ref{monoclinic}, 
as well as a visualization of the monoclinic angle $\beta$.
The introduction of a monoclinic angle $\beta$ causes a slight shear movement of the layers in the $ab$-plane. Lattice parameters and crystal symmetry 
at 300~K and 10~K taken at the SLS-MS beamline are listed in Table~\ref{Table1}. Refinements with space group $Cm$ describe the splitting of the Bragg 
peaks in the nuclear phase very well (see Fig.~\ref{D1A}(b)).
\begin{figure}
\includegraphics[width=\columnwidth]{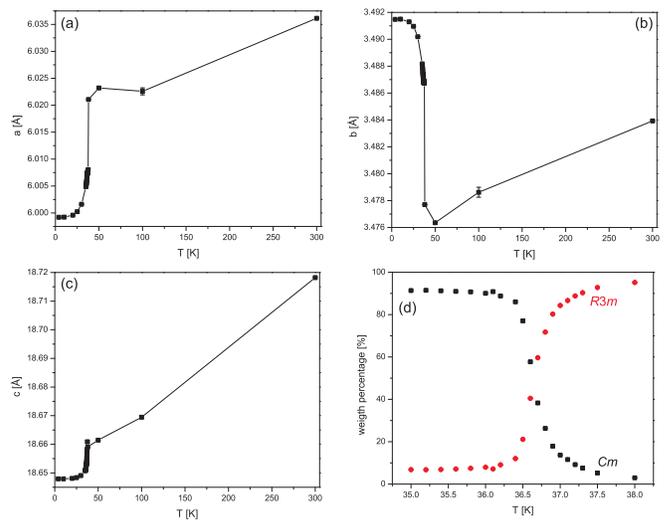}
\caption{(Color online) Temperature evolution of the lattice parameters a, b and c corresponding to (a) to (c), extracted from synchrotron data, in 
monoclinic description $Cm$. For better comparability the refinements for $T>37$~K were performed in the monoclinic description with 
$\beta=90^{\circ}$ and $a=\sqrt{3}\,b$. (d) shows the refined weight percentage of both the high temperature $R3m$ phase and the low temperature $Cm$ 
phase through the transition. The error bars are smaller than the symbols.}
\label{parameter}
\end{figure}
Above the phase transition all three lattice parameters, depicted in Figs.~\ref{parameter}(a) to \ref{parameter}(c), decrease with temperature. Upon 
further cooling $a$ and $c$ show a discontinuous jump towards lower values, whereas $b$ increases quickly. The complementarity of the in-plane lattice 
constants $a$ and $b$ will later be explained by a distortion of the Cr hexagon.
\begin{table}[h!]
\caption{Lattice parameters of CuCrS$_2$, refined with space group $R3m$ at 300~K and $Cm$ at 10~K. Data were collected at the SLS-MS beamline with 
wavelength $\lambda=0.621288(1)$~\AA.} 
\label{Table1}{\smallskip}
\begin{tabular}{c|cc}
 \hline
 Temperature & 300 K & 10 K\\
 \hline
 Space group & $R3m$ & $Cm$\\
 Nr.         &  160        &  8  \\
             & rhombohedral & monoclinic\\
 \hline
 $a$ [\AA]   & 3.484543(4)  & 5.99923(1)\\
 $b$ [\AA]   & 3.484543(4)  & 3.491502(5)\\
 $c$ [\AA]   & 18.71878(2)  & 18.64794(3)\\
 \hline
 $\alpha$ [$^{\circ}$]   & 90  & 90\\
 $\beta$ [$^{\circ}$]    & 90  & 89.93727(8)\\
 $\gamma$ [$^{\circ}$]   & 120 & 90\\
 \hline
 $R_{Bragg} [\%]$ & 4.13      & 3.83 \\
 \hline
\end{tabular} 
\end{table}\\
The refinement of magnetic intensities was done with a magnetic propagation vector documented earlier \cite{Wintenberger1987} which was redefined on 
the monoclinic lattice and fits well to the magnetic Bragg peaks (see Fig.~\ref{D1A}(a)). The propagation vector 
\mbox{\boldmath$k$}$=(-0.493,-0.087,1.25)$ in relative length units (r.l.u.), describes a three-dimensional helical arrangement of magnetic moments 
originating from Cr$^{3+}$ ions. The best fitting results were obtained with no magnetic moment on the Cu atoms, from which we conclude that Cu is in 
oxidation state Cu$^+$, corresponding to the stoichiometry of the sample. Neutron single crystal diffraction experiments were performed on IN3~(ILL) 
and TriCS (SINQ) to confirm the rather unusual magnetic propagation vector and to avoid artifacts from the powder refinement. Magnetic Bragg peaks 
were found at several positions in 3D reciprocal space, corresponding to the propagation vector obtained from powder measurements and verified the 
incommensurate magnetic structure. The intensity of all magnetic peaks, including the strongest (003)-\mbox{\boldmath$k$}, vanishes at $T_N=37.5$~K 
but shows different temperature behavior than in the powder, i.e. a varying critical exponent extracted from $M_s\sim\sqrt{I}$ close to the phase 
transition. The values are 0.08(4) for the powder compared to 0.5(1) for the single crystal. This is possibly due to an internal magnetic field in the 
single crystal caused by the ferromagnetic impurity CuCr$_2$S$_4$ which slows down the transition. The magnetic intensity is saturated for $T<30$~K. 
Since $T_N$ and \mbox{\boldmath$k$} coincide for powder and single crystal in both susceptibility and diffraction data, we conclude that the presence 
of CuCr$_2$S$_4$ does not disturb the measurement of the magnetic structure in CuCrS$_2$ at low temperatures.\\
A detailed analysis of the D1A powder data revealed an interesting coupling between magnetic ordering and structural distortion.
The temperature dependencies of the magnetic intensity in the strongest Bragg peak at $2\theta=13.96^{\circ}$ and the monoclinic angle $\beta$ show 
reverse behavior, depicted in Fig.~\ref{angle}.
As the sample is cooled below $T_N$ the magnetic ordering sets in and a deformation of the lattice develops simultaneously, expressed by a monoclinic 
angle $\beta$ deviating from 90$^{\circ}$. Although the deviation of $\beta$ from a higher symmetric value seems to be tiny it is crucial to solve the 
low temperature crystal structure. Figure~\ref{angle} shows the mutual influence of lattice distortion and magnetic ordering.\\
Additional to the interlayer shear movement, another type of intralayer structural distortion was identified from synchrotron powder measurements at 
the SLS-MS beamline. At high temperatures the in-plane chromium lattice describes a symmetric hexagon. Upon cooling the symmetric chromium environment 
distorts, such that four of the six equal Cr-Cr bonds contract and two elongate. This causes a flattening of the hexagon, as schematically shown in 
the inset of Fig. \ref{hexagon}, and leads to nearest-neighbor ($d_1$) and next nearest-neighbor ($d_2$) Cr-bonds. The Cr-Cr distances $d_1$ and $d_2$ 
show the same temperature evolution as the interlayer distortion (see Fig.~\ref{hexagon}). In addition, $d_2$ coincides with the lattice parameter 
$b$, which explains its unconventional temperature dependence below $T_N$.\\  
These two types of distortions, the interlayer shear movement by an angle $\beta$ and the intralayer hexagonal symmetry breaking, seem to play an 
important role for the magnetic ordering. The system is able to select a magnetic ground state due to the relief of geometrical frustration.

\section{Discussion}

The analysis of the diffraction pattern at different temperatures shows an abrupt change of Cr-Cu (lattice parameter $c$) and Cr-Cr distances ($d_2$) 
below the phase transition. Usually a contraction causes strain in the lattice which is proportional to the square of the atomic  displacement $u$ and 
increases elastic energy $\sim u^2$. A phase transition has to be energetically favorable, which means that the gain in elastic energy must be 
compensated. Using the example of the phase transition in solid oxygen, Rastelli et \textit{al.} \cite{Rastelli1988} state that a first-order 
transition is feasible when the ground-state energy of the rhombohedral phase equals the ground-state energy of the monoclinic phase. Synchrotron 
powder diffraction data confirm a first-order transition in CuCrS$_2$ identified from a residual rhombohedral $R3m$ phase even at the lowest 
temperature, depicted in Fig.~\ref{parameter} (d). Mixed-phase regimes are characteristic for first-order phase transitions \cite{Weissmantel1980}. 
This may explain the deviation of the critical exponent 0.08(4) from common models (Ising, XY or Heisenberg), since it is merely defined for 
second-order phase transitions. A reduction of the total free energy can only be achieved by a decrease in magnetic energy. For discussion we assume 
three different magnetic exchange paths in CuCrS$_2$ comprising in-plane nearest-neighbors $J_1$ (along $d_1$, see Fig.~\ref{hexagon}) and next 
nearest-neighbors $J_2$ (along $d_2$) and inter-plane exchange $J_3$.\\
The intralayer exchange integrals $J_1$ and $J_2$ can have two origins, a direct antiferromagnetic interaction of neighboring Cr$^{3+}$ ions or an 
indirect superexchange interaction via S$^{2-}$ which would be ferromagnetic due to the Cr-S-Cr angle of 90$^{\circ}$. The susceptibility curve shows 
antiferromagnetic behavior, and we can therefore neglect the ferromagnetic superexchange path. Interlayer exchange is presumably mediated by Cu$^+$ 
ions through the path Cr-S-Cu-S-Cr. It needs to be taken into account to explain the three-dimensional magnetic structure. The contraction of Cr-Cu 
bonds through the transition seems to provide evidence for the important role of the interlayer Cr-Cu-Cr exchange path. The ratio between the in-plane 
and inter-plane exchange constants can be calculated for a rhombohedral antiferromagnet with a helical propagation vector.\cite{Rastelli1988} For 
CuCrS$_2$ we find a ratio $J_3/J_1=2$, which would imply a very strong inter-plane coupling, but this value needs to be confirmed by inelastic neutron 
scattering.\\
The in-plane distortion of the high temperature symmetric Cr$^{3+}$-hexagon consists of a deformation where four nearest-neighbors move towards the 
central Cr$^{3+}$ ion and two nearest-neighbors move away. The transition takes place in order to maximize magnetic energy, as was found in solid 
oxygen \cite{Stephens1986}. The two different kinds of intralayer Cr-Cr distances $d_1$ and $d_2$ (Fig. \ref{hexagon}) may be evidence for the 
formation of magnetic clusters in CuCrS$_2$, which is supported by preliminary neutron scattering results. It has been found that localized magnetic 
excitations occur additionally to standard spin waves branches in the spectra. To our knowledge the formation of magnetic clusters in triangular 
antiferromagnets with spin-lattice coupling has only been theoretically described \cite{Jia2006} so far. This phenomenon makes CuCrS$_2$ an 
interesting candidate for studying new magnetic effects on frustrated lattices.

\section{Conclusion}

From the quasi two-dimensional layered structure of CuCrS$_2$ one would expect a low dimensional magnetic ordering due to the van der Waals gap 
between magnetic Cr$^{3+}$ layers. Nevertheless, our measurements show unambiguously that magnetic moments in CuCrS$_2$ order in a fully 
three-dimensional manner with a helical magnetic propagation vector \mbox{\boldmath$k$}$=(-0.493,-0.087,1.25)$.  Moreover, the system undergoes a 
crystallographic transition from rhombohedral $R3m$ to monoclinic $Cm$ at $T_N=37.5$~K, which coincides with susceptibility measurements. Since the 
ordering of magnetic moments and the lattice distortion occur simultaneously, CuCrS$_2$ has been identified as magnetoelastic material. The system is 
therefore an interesting candidate for studying spin-lattice effects on a triangular lattice.

\section{Acknowledgment}

We are grateful for support and allocated beam time at the Institut Laue-Langevin (D1A, IN3), Grenoble, France, the spallation neutron source SINQ 
(TriCS, DMC) and the SLS-MS beamline, both Paul Scherrer Insitut, Villigen, Switzerland. This work was supported by INTAS grant 06-1000013-9002 of the 
Russian Academy of Science (RAS), Siberian Branch.

\bibliographystyle{uuu}
{}

\end{document}